\documentclass[reprint,9pt]{sigplanconf}

\usepackage{amsmath}

\begin{document}

\conferenceinfo{ILC'09}{March 22--25, 2009, Cambridge, Massachusetts, USA}
\copyrightyear{2009}
\copyrightdata{}
\proceedings{Proceedings of the International Lisp Conference}

\toappear{}

\titlebanner{DRAFT---Do not distribute}
\preprintfooter{Chun Tian (binghe)'s paper for ILC 2009}

\title{SNMP for Common Lisp}

\authorinfo{Chun Tian (binghe)}
           {Hangzhou Research Center, P. R. China\\NetEase.com, Inc.}
           {binghe.lisp@gmail.com}

\maketitle

\begin{abstract}
  Simple Network Management Protocol (SNMP) is widely used for management of
  Internet-based network today. In Lisp community, there're large Lisp-based
  applications which may need be monitored, and there're Lispers who may need
  to monitor other remote systems which are either Lisp-based or not.
  However, the relationship between Lisp and SNMP haven't been studied enough
  during past 20 years.
  
  The
  \textsc{cl-net-snmp}\footnote{\texttt{http://common-lisp.net/project/cl-net-snmp}}
  project has developed a new Common Lisp package which implemented the
  SNMP protocol. On client side, it can be used to query remote SNMP peers,
  and on server side, it brings SNMP capability into Common Lisp based applications,
  which could be monitored from remote through any SNMP-based management system.
  It's also a flexible platform for researches on network management and SNMP itself.
  But the most important, this project tries to prove: \textbf{Common Lisp is the most
  fit language to implement SNMP}.

  Different from other exist SNMP projects on Common Lisp,
  \textsc{cl-net-snmp} is clearly targeted on full SNMP
  protocol support include \textsl{SNMPv3} and server-side work (agent). During
  the development, an general ASN.1 compiler and runtime package and
  an portable UDP networking
  package are also implemented, which would be useful for other related
  projects.

  In this paper, the author first introduces the SNMP protocol and
  a quick tutorial of \textsc{cl-net-snmp}
  on both client and server sides, and then the Lisp native design
  and the implementation details of the ASN.1 and SNMP package,
  especially the ``code generation'' approach on compiling SNMP MIB
  definitions from ASN.1 into Common Lisp.
\end{abstract}

\category{C.2.2}{Computer-Communication Networks}{Network Protocols}[Applications]
\category{C.2.3}{Computer-Communication Networks}{Network Operations}[Network monitoring]
\category{C.2.6}{Computer-Communication Networks}{Internetworking}[Standards]
\category{D.3.4}{Programming Languages}{Processors}[Code generation]

\terms
Languages, Network Management

\keywords
Lisp, SNMP, ASN.1

\section{SNMP Overview}

Simple Network Management Protocol (SNMP) is the \textit{de facto}
standard for network management and service monitoring. Using SNMP,
people can monitor the status of remote UNIX servers and network
equipment.

Though there're other protocols defined before and after
SNMP, it has several advantages which lead it become popular. First,
from the view of implementation costs, SNMP is the most lightweight and
can be easily implemented by hardware. Second, it's really simple to
do the actual management job: all information are defined as variables
which are either scalars or conceptually organized into tables. All of
these variables are formally defined by using a definition language
called the Structure of Management Information (SMI) \cite{RFC:2578},
which is a subset
of Abstract Syntax Notation One (ASN.1) \cite{ISO:ASN.1}.
Data definitions written
in the SMI are called Management Information Base (MIB) \cite{RFC:3418}.
Third, SNMP has minimum resource requirements.
By using User Datagram Protocol (UDP) \cite{RFC:768},
most SNMP operations only need one pair of UDP packets: a request and
a response.

SNMP has three major versions: \textsl{SNMPv1} \cite{RFC:1157},
\textsl{SNMPv2c} \cite{RFC:1901} and
\textsl{SNMPv3} \cite{RFC:3411}. The differences between \textsl{SNMPv1} and
\textsl{SNMPv2c} is mainly on SMI and MIB side, no packet format changes.
The \textsl{SNMPv3} is a big step towards security: it supports authentication
and encryption based on standard algorithms, and the packet format
changed a lot.

The relationship between SNMP and ASN.1 is very important because
any implementation of SNMP must first implement ASN.1, at least a subset of it.
As we already mentioned above, the MIB are defined by a subset of ASN.1,
the SMI. Actually, the SNMP message format are just defined through
ASN.1: The whole SNMP message is type of ASN.1 \texttt{SEQUENCE}, and
all SNMP protocol data units (PDU) are defined in framework of ASN.1 type
system. ASN.1 data is just abstract objects, it need to be translated
into octet bytes and then back. This translation is generally called encoding and
decoding.
The encoding/decoding method chosen by SNMP is the
Basic Encoding Rule (BER) \cite{ISO:BER},
which use a Type-Length-Value (TLV) combination in representation of any
ASN.1 data as octet bytes in networking packets.

There're many SNMP-related open source and commercial software in the
world. On \textsl{SourceForge.net}, there're almost 300 projects
matching the keyword ``SNMP'', which can be classified into three
types:
\begin{itemize}
\item SNMP library or extension for specific programming language.
\item SNMP agent which acts as a standalone agent or agent extension.
\item SNMP manager which is used for managing SNMP-enabled servers or
  equipment, either GUI or Web-based.
\end{itemize}

From the view of programming languages, almost every in-use language
has a well implemented SNMP library/package. For popular languages
like Java, Python or C\#, there're often several similar projects
existing in competition. And there's at least one language, Erlang,
which ships full SNMP support (more than 60,000 lines of code) with
the language itself \footnote{\texttt{http://erlang.org/}}.
J. Schonwalder wrote a good summary \cite{Schonwalder2002}
which mentioned various open source SNMP library/tools for different languages.

On the other side, the relationship between Common Lisp and ASN.1/SNMP
haven't been studied enough before. There're some similarities between
ASN.1 and Common Lisp. First, the most important ASN.1 aggregate
type \texttt{SEQUENCE} can be directly mapped to Common Lisp lists,
the both are generalized lists which could contain any object. Second,
ASN.1 lexical tokens like multi-line strings, comments, and the number
representation are almost the same as those in Common Lisp. Later will show,
by making some necessary changes to CL readtables, the Lisp reader
could directly read ASN.1 definitions into Lisp tokens. For SNMP itself,
there're also some connection between its design and Lisp, especially the
design of ``GetNextRequestPDU'' \cite{RFC:3416},
it's just like the Common Lisp function
\texttt{cdr} and later will show that the implementation of this PDU has
almost just used the function \texttt{cdr}.

For Common Lisp, before 2007, two SNMP projects have been in
existence. Simon Leinen's
\textsc{sysman}\footnote{\texttt{http://www.switch.ch/misc/leinen/snmp/sysman.html}}
is the first Lisp-SNMP solution which supports client side
\textsl{SNMPv1}/\textsl{SNMPv2c} and server side agent on
Symbolics Lisp Machine. It's a pure lisp implementation, but
is not being maintained anymore, and doesn't support the new protocol \textsl{SNMPv3}.
During 2006, Larry Valkama wrote two SNMP client packages
\footnote{\texttt{http://www.valkama.se/article/lisp/snmp}},
both of which are not pure lisp:
one is a FFI wrapper to Net-SNMP library, and the other works
by capturing outputs of Net-SNMP's common-line utilities.

Unfortunately neither of above projects had showed the advantages of using
Common Lisp to implement SNMP. There're two ways to implement the SNMP protocol:
with and without MIB support, one is hard and the other is simple. The simple
way to implement SNMP is just use number lists as ASN.1 \texttt{OBJECT IDENTIFIER} (OID)
such as ``1.3.6.1.4.1.1.1.0'', and doesn't care their additional type information
beside ASN.1 itself, the
\textsc{snmp1}\footnote{\texttt{http://common-lisp.net/projects/snmp1}} is
a example, it just implement the BER encoding and SNMP packets
generation, no matter what the OID numbers mean.

The ``hard'' way to implement SNMP, the relation between OID number lists
and their names must be kept, and these information should be retrieve from the
original MIB definitions. To achieve this, an ASN.1 syntax parser would be needed,
and the structure of a MIB storage system should be designed well.

The \textsc{cl-net-snmp} project solved all above issues well and went
``a hardest way'' to implement SNMP. 1) It have an ASN.1 to Common Lisp
compiler which compiles MIB definitions into Common Lisp source code which
then defines almost all information used by SNMP package. 2) The CLOS-based BER
encoding/decoding system in ASN.1 package is extensible: user can define their
own new ASN.1 types, and all SNMP PDUs are defined in this way. 3) The SNMP
package support full MIB, that means all information defined in
MIB are being used when doing SNMP work. 4) Object-orient SNMP query facility.
\textsc{cl-net-snmp} runs on the following Common Lisp
implementations: CMUCL, SBCL, Clozure CL, LispWorks, Allegro CL and
Scieneer CL; and runs under Linux, Solaris, Mac OS X and Windows.

Following sections will first introduce the SNMP package from user view
and then show the design idea and implementation details behind
the express and convenient API.

\section{\textsc{cl-net-snmp} Tutorial}

\subsection{Client-side SNMP}

The client-side API of SNMP package is quite straight-forward. The
central object which operated by almost all client functions is the
``SNMP session''. To query a remote SNMP peer, a session object should
be created first.

As SNMP protocol has three versions (\textsl{SNMPv1}, \textsl{SNMPv2c}
and \textsl{SNMPv3}), correspondingly, we have three session classes:
\texttt{v1-session}, \texttt{v2c-session} and \texttt{v3-session}. The
entry API of client-side SNMP is the function
\texttt{snmp:open-session}, which creates a new SNMP session:
\begin{verbatim}
snmp:open-session (host &key port version community
                             user auth priv)
\end{verbatim}

\subsubsection{SNMPv1 and SNMPv2c}

To create a \texttt{SNMPv1} or \texttt{SNMPv2c} session, only keywords \texttt{port},
\texttt{version} and \texttt{community} are needed. Suppose we have a
SNMP server whose host name is \texttt{"binghe-debian.local"},
which is running a Net-SNMP agent on default port 161, its SNMP
community is \texttt{"public"}, and the SNMP protocol is
\textsl{SNMPv2c}, then the following form will create a new session
and assign it to a variable \texttt{s1}:
\begin{verbatim}
> (setf s1 (snmp:open-session "binghe-debian.local"
                              :port 161
                              :version :v2c
                              :community "public"))
#<SNMP::V2C-SESSION 223CF317>
\end{verbatim}

In current version of SNMP package, when a session is being created, a
new socket will be opened at the same time. You can use
\texttt{snmp:close-session} to close the session:
\begin{verbatim}
snmp:close-session (session)
\end{verbatim}

All SNMP PDUs \cite{RFC:3416} are supported. When a session is opened, functions which
can be used on it are listed below:

\begin{itemize}
\item \texttt{snmp:snmp-get}
\item \texttt{snmp:snmp-get-next}
\item \texttt{snmp:snmp-walk}\footnote{\texttt{snmp:snmp-walk} is a compound operation,
it may calls \texttt{snmp:snmp-get} and \texttt{snmp:snmp-get-next} to do
the actual work.}
\item \texttt{snmp:snmp-set}
\item \texttt{snmp:snmp-trap}\footnote{\texttt{snmp:snmp-trap} only defined in \textsl{SNMPv1}}
\item \texttt{snmp:snmp-inform}
\item \texttt{snmp:snmp-bulk}
\end{itemize}

For normal lisp applications, \texttt{snmp:snmp-get} is the most
useful function. Users can retrieve multiple variables in one query as
the SNMP protocol supported:
\begin{verbatim}
> (snmp:snmp-get s1 '("sysDescr.0" "sysName.0"))
("Linux binghe-debian.local 2.6.26-1-amd64 #1
  SMP Thu Oct 9 14:16:53 UTC 2008 x86_64"
 "binghe-debian.local")
\end{verbatim}
While only one variable is queried, \texttt{snmp:snmp-get} can be used
just like this:
\begin{verbatim}
> (snmp:snmp-get s1 "sysName.0")
"binghe-debian.local"
\end{verbatim}

The string \texttt{"sysDescr.0"} here will be translated to a
ASN.1 OID instance. When the SNMP client operated on
multiple servers, preparing all OID instances before the actual query work
would increase the performance. The function \texttt{asn.1:oid} is used for
this translation:
\begin{verbatim}
> (asn.1:oid "sysName.0")
#<ASN.1:OBJECT-ID SNMPv2-MIB::sysName.0>
> (snmp:snmp-get s1 *)
"binghe-debian.local"
\end{verbatim}

The \texttt{snmp:with-open-session} macro can be used to establish a
temporary session:
\begin{verbatim}
with-open-session ((session &rest args) &body body)
\end{verbatim}
Following is a sample query:
\begin{verbatim}
> (snmp:with-open-session (s "binghe-debian.local"
                             :port 161
                             :version :v2c
                             :community "public")
    (snmp:snmp-get s '("sysName.0")))
("binghe-debian.local")
\end{verbatim}
Actually, the SNMP port as 161, community as ``public'' and version as
\textsl{SNMPv2c} are default settings, which have been held by three
Lisp variables:
\begin{verbatim}
(in-package :snmp)

(defvar *default-snmp-version* +snmp-version-2c+)
(defvar *default-snmp-port* 161)
(defvar *default-snmp-community* "public")
\end{verbatim}

When operating on default settings, the query syntax can also be
simplified into a hostname string instead of SNMP session instance:
\begin{verbatim}
> (snmp:snmp-get "binghe-debian.local" "sysName.0")
"binghe-debian.local"
\end{verbatim}

\subsubsection{SNMPv3}

The major visibility changes of \textsl{SNMPv3} \cite{RFC:3411} are authenticate
and encryption support.
Opening an \textsl{SNMPv3} session needs more different keywords
besides \texttt{host} and \texttt{port}:

\begin{itemize}
\item \texttt{version}, possible values for \textsl{SNMPv3} are
  keyword \texttt{:v3}, \texttt{:version-3} and constant
  \texttt{snmp:+snmp-version-3+}.

\item \texttt{user}: A string as the SNMP security name \cite{RFC:3414}.

\item \texttt{auth}: Authenticate protocol and key, valid arguments:
  $\langle\mathrm{string}\rangle$,
  \texttt{($\langle\mathrm{string}\rangle$
    $\langle\mathrm{protocol}\rangle$)} or
  \texttt{($\langle\mathrm{string}\rangle$
    . $\langle\mathrm{protocol}\rangle$)}, which the
  \texttt{$\langle\mathrm{protocol}\rangle$} can be \texttt{:md5}
  (default) or \texttt{:sha1}.

\item \texttt{priv}: Encryption/privacy protocol and key, valid
  arguments: $\langle\mathrm{string}\rangle$,
  \texttt{($\langle\mathrm{string}\rangle$
    $\langle\mathrm{protocol}\rangle$)} or
  \texttt{($\langle\mathrm{string}\rangle$
    . $\langle\mathrm{protocol}\rangle$)}, which the
  \texttt{$\langle\mathrm{protocol}\rangle$} can only be \texttt{:des}
  at this time.
\end{itemize}

When both \texttt{auth} and \texttt{priv} are \texttt{nil},
\textsl{SNMPv3} operates at security level ``noAuthNoPriv''; when only
\texttt{auth} is set up, the security level is ``authNoPriv''; and
when both are set up, the strongest method ``authPriv'' is used. When
\textsl{SNMPv3} is being used, all arguments must be set explicitly by
\texttt{snmp:open-session} or \texttt{snmp:with-open-session}. There's
no express way as those in earlier SNMP protocol versions.

For example, assume we have a remote SNMP peer which works through the
following specification:
\begin{itemize}
\item \texttt{(user "readonly")} (Security name is ``readonly'')
\item \texttt{(auth (:md5 "ABCDEFGHABCDEFGH"))} (Authenticate protocol
  is MD5, followed by the authenticate key)
\end{itemize}

Then a quick query on ``sysDescr.0'' would be:
\begin{verbatim}
> (snmp:with-open-session
      (s "binghe-debian.local"
         :version :v3 :user "readonly"
         :auth '(:md5 "ABCDEFGHABCDEFGH"))
    (snmp:snmp-get s "sysDescr.0"))
"Linux binghe-debian.local 2.6.26-1-amd64 #1
 SMP Thu Oct 9 14:16:53 UTC 2008 x86_64"
\end{verbatim}

Here the \texttt{auth} argument \texttt{(:md5 "ABCDEFGHABCDEFGH")} can
also use just \texttt{"ABCDEFGHABCDEFGH"} instead. That's because
MD5\footnote{The actual authenticate protocol used by SNMP is
  \textsl{HMAC-MD5-96} and \textsl{HMAC-SHA1-96}.} is the default
authenticate protocol.

\subsubsection{High-level SNMP Query}

Recently a new feature has been added to \textsc{cl-net-snmp}: object-oriented
and SQL-like SNMP query. The idea of using SQL-like syntax on query SNMP
variables can be traced back to Wengyik Yeong's paper \cite{SNMPql} in 1990.
However, in near 20 years, there's no other implementation on this idea.
By using Common Lisp, the most dynamic programming language, \textsc{cl-net-snmp}
goes even further.

Query technologies is mainly used on \textsl{selective} or
\textsl{fast} information retrieval from MIB tables.
In \textsc{cl-net-snmp}, the ASN.1 compiler can compile MIB definitions
into Common Lisp code. During this process, not only OID name to numbers
information are saved, but also the structure of MIB tables. For example,
the MIB table ``ifTable'' which contains network interface information,
it's mainly defined by following MIB:
\begin{verbatim}
ifTable OBJECT-TYPE
    SYNTAX      SEQUENCE OF IfEntry
    MAX-ACCESS  not-accessible
    STATUS      current
    DESCRIPTION
        "A list of interface entries.  The number of
         entries is given by the value of ifNumber."
    ::= { interfaces 2 }
\end{verbatim}
And the ``IfEntry'' is a ASN.1 type of \texttt{SEQUENCE}:
\begin{verbatim}
IfEntry ::=
    SEQUENCE {
        ifIndex           InterfaceIndex,
        ifDescr           DisplayString,
        ifType            IANAifType,
        ifMtu             Integer32,
        ifSpeed           Gauge32,
        ifPhysAddress     PhysAddress,
        ifAdminStatus     INTEGER,
        ifOperStatus      INTEGER,
        ifLastChange      TimeTicks,
        ...
        ifSpecific        OBJECT IDENTIFIER
    }
\end{verbatim}
The ASN.1 compiler can compile above ``IfEntry'' type into a CLOS class definition:
\begin{verbatim}
(defclass |IfEntry| (sequence-type)
  ((|ifIndex| :type |InterfaceIndex|)
   (|ifDescr| :type |DisplayString|)
   (|ifType| :type |IANAifType|)
   (|ifMtu| :type |Integer32|)
   (|ifSpeed| :type |Gauge32|)
   (|ifPhysAddress| :type |PhysAddress|)
   (|ifAdminStatus| :type integer)
   (|ifOperStatus| :type integer)
   (|ifLastChange| :type |TimeTicks|)
   ...
   (|ifSpecific| :type object-id)))
\end{verbatim}
These code could be compiled and loaded with SNMP package, and
dynamic loading MIB files should be possible: most Common Lisp platform
supports defining new CLOS classes even after delivery.
Once the structure of MIB tables are known, rest work will be quite easy.
A query on all interfaces information is like this:
\begin{verbatim}
> (snmp:snmp-select "ifTable"
                    :from "binghe-debian.local")
(#<ASN.1/IF-MIB::|IfEntry| 200E8A43>
 #<ASN.1/IF-MIB::|IfEntry| 200A5DDF>)
\end{verbatim}
\texttt{snmp:snmp-select} is a high-level API. It could return CLOS instances
instead of lists, and a simple query like above may involve various low-level
SNMP operations. The OO idea is learnt from LispWorks
\textsl{CommonSQL}.
The internal operations of above query will first use
\texttt{snmp:snmp-get-next} to test how many ``lines'' does the table have.
In this example, it has just two lines. Once the number of lines is known,
then just using \texttt{snmp:snmp-get} to get each line will be OK. In above
query example on \texttt{SNMPv2c}, there're only \textbf{4} UDP packets sent
to get the whole table. Compare to that, the tradition way
by using \texttt{snmp:snmp-walk} will costs 44 UDP packets (the column length
of this table times its lines).

There're two ways to get the actual data in returning instances. Assumes
the first instance has been stored in a variable \texttt{interface}. One
way is using \texttt{asn.1:plain-value} to convert the instance into lists
of values:
\begin{verbatim}
> interface
#<ASN.1/IF-MIB::|IfEntry| 200A170B>

> (asn.1:plain-value interface)
((#<ASN.1:OBJECT-ID IF-MIB::ifIndex (1) [0]> 2)
 (#<ASN.1:OBJECT-ID IF-MIB::ifDescr (2) [0]> "eth0")
 (#<ASN.1:OBJECT-ID IF-MIB::ifType (3) [0]> 6)
 (#<ASN.1:OBJECT-ID IF-MIB::ifMtu (4) [0]> 1500)
 (#<ASN.1:OBJECT-ID IF-MIB::ifSpeed (5) [0]>
  #<ASN.1:GAUGE 10000000>)
 ...
 (#<ASN.1:OBJECT-ID IF-MIB::ifSpecific (22) [0]>
  #<ASN.1:OBJECT-ID SNMPv2-SMI::zeroDotZero (0) [0]>))
\end{verbatim}
To retrieve specific item, the funtion \texttt{asn.1:slot-value-using-oid}
can be used:
\begin{verbatim}
> (asn.1:slot-value-using-oid interface "ifDescr")
"eth0"
\end{verbatim}

Conditional query is still under research. The most native way to
represent SQL \texttt{WHERE} clause in \texttt{snmp:snmp-select}
haven't determined.

\subsection{Server-side SNMP}

Server-side SNMP is mainly used for Lisp image being queried from
outside. The entry API is how to start and stop the SNMP server, this
can be done by \texttt{snmp:enable-snmp-service} and
\texttt{snmp:disable-snmp-service}:
\begin{verbatim}
> (snmp:enable-snmp-service)
#<SNMP:SNMP-SERVER SNMP Server at 0.0.0.0:8161>
\end{verbatim}
Above function will open a new Lisp thread which acts as a SNMP
server. By default, the SNMP server listens on port 8161 and
wild-cast (\texttt{0.0.0.0}) address. At current stage, no access
control\footnote{The access control protocol used by SNMP is called View-based Access Control Model (VACM) \cite{RFC:3415}.}
is implemented and only \textsl{SNMPv1/SNMPv2c} are
supported. We can use the SNMP client API to query it:
\begin{verbatim}
> (setf snmp:*default-snmp-port* 8161)
8161

> (snmp:snmp-get "localhost" "sysDescr.0")
"LispWorks Personal Edition 5.1.1 on
 binghe-mac.people.163.org"
\end{verbatim}
Here we changed the default SNMP port to make things easier. Or we can
use command-line utilities from Net-SNMP project:

\begin{verbatim}
$ snmpget -v 2c -c public localhost:8161 sysDescr.0
SNMPv2-MIB::sysDescr.0 = STRING: LispWorks Personal\
 Edition 5.1.1 on binghe-mac.people.163.org
\end{verbatim}

This time we do a ``SNMP walk'' on MIB ``system'' node:
\begin{verbatim}
> (snmp:snmp-walk "localhost" "system")
((#<ASN.1:OBJECT-ID SNMPv2-MIB::sysDescr.0>
  "LispWorks Personal Edition 5.1.1 on
   binghe-mac.people.163.org")
 (#<ASN.1:OBJECT-ID SNMPv2-MIB::sysObjectID.0>
  #<ASN.1:OBJECT-ID
    LISP-MIB::clNetSnmpAgentLispWorks (5) [0]>)
 (#<ASN.1:OBJECT-ID
    DISMAN-EVENT-MIB::sysUpTimeInstance.0>
  #<ASN.1:TIMETICKS (69652) 0:11:36.52>)
 (#<ASN.1:OBJECT-ID SNMPv2-MIB::sysContact.0>
  "Chun Tian (binghe) <binghe.lisp@gmail.com>")
 (#<ASN.1:OBJECT-ID SNMPv2-MIB::sysName.0>
  "binghe-mac.local (binghe-mac)")
 (#<ASN.1:OBJECT-ID SNMPv2-MIB::sysLocation.0>
  "binghe-mac.people.163.org")
 ...)
\end{verbatim}

An SNMP server is only useful when there is useful information in it.
It's extensible: new SNMP scalar variables or tables can be defined on the
fly. There're two high-level API macros which can be used:
\texttt{snmp:def-scalar-variable} and \texttt{snmp:def-listy-mib-table}.
\begin{verbatim}
def-scalar-variable (name (agent) &body body)
def-listy-mib-table (name (agent ids) &body body)
\end{verbatim}
And a low-level API function \texttt{snmp:register-variable}:
\begin{verbatim}
register-variable (oid function &key dispatch-table
                                walk-table walk-list)
\end{verbatim}
For example, the variable ``sysDescr.0'' is defined by the following
form:
\begin{verbatim}
(def-scalar-variable "sysDescr" (agent)
  (format nil "~A ~A on ~A"
          (lisp-implementation-type)
          (lisp-implementation-version)
          (machine-instance)))
\end{verbatim}
When the SNMP server running in LispWorks being queried from outside
by Net-SNMP utilities, the query to ``sysDescr.0'' may shows:
\begin{verbatim}
$ snmpget -v 2c -c public binghe-debian.local:8161\
  sysDescr.0

SNMPv2-MIB::sysDescr.0 = STRING:\
 LispWorks 5.1.1 on binghe-debian.local
\end{verbatim}
The \texttt{agent} parameter in above
\texttt{snmp:def-scalar-variable} form is used to refer to current
SNMP agent instance, which can be used to access the status of current
agent. For example:
\begin{verbatim}
(def-scalar-variable "snmpInPkts" (agent)
  (counter32 (slot-value agent 'in-pkts)))
\end{verbatim}
When ``snmpInPkts.0'' is queried, the slot value \texttt{in-pkts} of current
agent instance will be coerced into ASN.1 \texttt{Counter32} type and
returned.\footnote{This variable has't been actually used in current version.}

A MIB Table can be defined as per column. For example, to define a
column from ``sysORUpTime.1'' to ``sysORUpTime.9'', we have three
ways:
\begin{verbatim}
(def-listy-mib-table "sysORUpTime" (agent ids)
  (if (null ids)
    '((1) (2) (3) (4) (5) (6) (7) (8) (9))
    (timeticks 0)))

(def-listy-mib-table "sysORUpTime" (agent ids)
  (if (null ids)
    '(1 2 3 4 5 6 7 8 9)
    (timeticks 0)))

(def-listy-mib-table "sysORUpTime" (agent ids)
  (if (null ids)
    9
    (timeticks 0)))
\end{verbatim}

The \texttt{ids} parameter means the \textbf{rest} OID components
when a query gets to current \textbf{base} OID (``sysORUpTime'' here).
You can treat the \texttt{snmp:def-listy-mib-table} body as an ordinary
function body. When
it's called by an \texttt{ids} argument as \texttt{nil}, it should
return all its valid children to the SNMP agent. A list \texttt{'((1)
  (2) (3) (4) (5) (6) (7) (8) (9))} means each valid rest OID
has only one element. For the first child, it's just the number
``1''. When a child has rest OID as only one element, the
long list can be simplified into \texttt{'(1 2 3 4 5 6 7 8 9)}, which
just means ``sysORUpTime.1'', ``sysORUpTime.2'',
... ``sysORUpTime.9''. And, when all the children have a single number
and they are consistent (from 1 to N), the list can be again simplified
into just one number ``9'' which also means from 1 to 9, then
\texttt{(1)} to \texttt{(9)}.

For dynamic MIB tables, just let the form (as a function) returns an
dynamic list when \texttt{ids} given \texttt{nil} will work, like this
one:
\begin{verbatim}
(def-listy-mib-table "lispFeatureName" (agent ids)
  (let* ((features *features*)
         (number-of-features
          (list-length features)))
    (if (null ids)
      number-of-features
      (when (plusp (car ids))
        (->string (nth (mod (1- (car ids))
                            number-of-features)
                       features))))))
\end{verbatim}

Above ``lispFeatureName'' can return all elements of
\texttt{*features*} in current Lisp system as ASN.1 \texttt{OCTET
  STRING}. Every time it's called, the \texttt{number-of-features}
will be calculated. It's not 	quite optimized here. If you want faster
reply, the list count progress should be defined outside of the
function and as a cache to the actual value, and a separate thread may
be used to update all these parameter values on schedule.

To use above two macros for user-defined MIB nodes, named OID nodes
must be defined first, like above ``sysDescr'' or
``sysORUpTime''. There are two ways to achieve the goal: define a
MIB file in ASN.1 syntax and use the ASN.1 compiler from ASN.1 package to
translate it into LISP source code, or
directly write the Lisp version of the MIB definition:

\begin{verbatim}
(defoid |lispFeatureName| (|lispFeatureEntry| 2)
  (:type 'object-type)
  (:syntax '|DisplayString|)
  (:max-access '|read-only|)
  (:status '|current|)
  (:description
   "The string name of each element in *features*."))
\end{verbatim}

At compile-time, above definitions then will be translated into following form:
\begin{verbatim}
(IN-PACKAGE "ASN.1/LISP-MIB")

(PROGN
  (DEFVAR |lispFeatureName|
    (OR (ENSURE-OID |lispFeatureEntry| 2)
        (MAKE-INSTANCE 'OBJECT-ID
          :NAME '|lispFeatureName|
          :VALUE 2
          :PARENT |lispFeatureEntry|
          :MODULE *CURRENT-MODULE*
          :TYPE 'OBJECT-TYPE
          :SYNTAX '|DisplayString|
          :MAX-ACCESS '|read-only|
          :STATUS '|current|
          :DESCRIPTION
  "The string name of each element in *features*.")))
  (EVAL-WHEN (:LOAD-TOPLEVEL :EXECUTE)
    (ASN.1::REGISTER-OID
      (SYMBOL-NAME '|lispFeatureName|)
      '|lispFeatureName|)))
\end{verbatim}

The low-level function \texttt{snmp:register-variable} is used by
\texttt{snmp:def-scalar-variable} and
\texttt{snmp:def-listy-mib-table}.  Above definitions for ``sysDescr''
will be macro-expanded into:
\begin{verbatim}
(IN-PACKAGE "SNMP")

(PROGN
  (DEFUN |ASN.1/SNMPv2-MIB|::|sysDescr|
         (AGENT &OPTIONAL #:G1290)
    (DECLARE (IGNORABLE AGENT))
    (IF (NULL #:G1290)
        0
      (FORMAT NIL
              "~A ~A on ~A"
              (LISP-IMPLEMENTATION-TYPE)
              (LISP-IMPLEMENTATION-VERSION)
              (MACHINE-INSTANCE))))
  (EVAL-WHEN (:LOAD-TOPLEVEL :EXECUTE)
    (REGISTER-VARIABLE
      (OID "sysDescr")
      #'|ASN.1/SNMPv2-MIB|::|sysDescr|)))
\end{verbatim}

Besides, \texttt{snmp:register-variable} has some additional keywords, which
can be used explicitly to define MIB nodes in SNMP agent other than
the default. It's possible to run multiple different SNMP agents
simultaneously, but more codes are needed. You even cannot use
\texttt{snmp:enable-snmp-service} here.

\section{LISP-MIB}

Though just registering an OID node or sub-tree in SNMP server will
fit the goal for querying information from remote SNMP peers, if there's
no coordination on places, conflict would happen and information
defined by different SNMP vendors would be impossible to live
together. SNMP community spend so much time on how to define a common
framework to hold all variables from SNMP
vendors, that is the MIB (Management Information Base)\cite{RFC:3418}.

Another side: a SNMP server running in Lisp image should be possible
to reply the status of Lisp system itself, for example, the most
basically, implementation type and version, which can be returned by
standard functions. ANSI Common Lisp has also defined some status
functions of the Lisp system itself (i.e. the internal run time and
real time), useful constants
(i.e. \texttt{internal\-time-\-units-\-per\--second}), and special variables
(\texttt{*read-\-eval*}, \texttt{*print-\-circle*}, ...) All these
information plus implementation-specific data maybe useful to monitor
and just query from outside world. For popular Lisp packages and some
small applications which have their own status and parameters, the
requirement for MIB sub-tree should also be considered.

There's one place in MIB tree which just is left for SNMP vendors: the
``enterprises''
node\footnote{\textsl{iso.org.dod.internet.private.enterprises} (OID:
  1.3.6.1.4.1)}. Since there's no Lisp-related MIB registered before,
A new enterprise number from IANA
\footnote{\texttt{http://www.iana.org/assignments/enterprise-numbers}}
has been registered: \textbf{31609 (lisp)},
which allocated to the \textbf{LISP-MIB}.

The root of LISP-MIB is ``enterprises.lisp'' (31609). Its two children
are ``common-lisp'' and ``scheme''.

In ``common-lisp'' node, there're four common children at present:
\begin{itemize}
\item \texttt{lispSystem}, the summary information of current Lisp
  system.
\item \texttt{lispConstants}, constants of limits of number-types.
\item \texttt{lispPackages}, information store for lisp packages
  (utilities).
\item \texttt{lispApplications}, information store for lisp
  applications.
\end{itemize}
Other children of ``common-lisp'' node are reserved for Common Lisp
implementations.  Implementation-specific variables should be put
there.

The framework of LISP-MIB\footnote{The MIB definition of
  LISP-MIB and other LISP-*-MIBs in ASN.1 can be found in \textsc{cl-net-snmp}'s
  Subversion repository:
  \texttt{https://cl-net-snmp.svn.sourceforge.net/svnroot/cl-net\-snmp/snmp/trunk/asn1/lisp}}
is shown in Figure \ref{fig:lisp-mib}.

\begin{figure}
\begin{verbatim}
lisp (31609)
  common-lisp (1)
    lispSystem (1)
      lispImplementationType (1)
      lispImplementationVersion (2)
      lispLongSiteName (3)
      lispShortSiteName (4)
      lispMachineInstance (5)
      lispMachineType (6)
      lispMachineVersion (7)
      lispSoftwareType (8)
      lispSoftwareVersion (9)
      lispInternalRealTime (10)
      lispInternalRunTime (11)
      lispInternalTimeUnitsPerSecond (12)
      lispUniversalTime (13)
      lispFeatureTable (14)
      lispPackageTable (15)
      lispModuleTable (16)
    lispConstants (2)
      lispMostPositiveShortFloat (1)
      lispLeastPositiveShortFloat (2)
      lispLeastPositiveNormalizedShortFloat (3)
      ...
    lispPackages (3)
      cl-net-snmp (1)
        clNetSnmpObjects (1)
        clNetSnmpEnumerations (2)
          clNetSnmpAgentOIDs (1)
            clNetSnmpAgent (1)
            clNetSnmpAgentLispWorks (5)
            clNetSnmpAgentCMUCL (6)
            ...
      cl-http (2)
      ...
    lispApplications (4)
    lispworks (5)
    cmucl (6)
    sbcl (7)
    clozure (8)
    allegro (9)
    scl (10)
    ...
  scheme (2)
\end{verbatim}  
  \caption{LISP-MIB}
  \label{fig:lisp-mib}
\end{figure}

\section{Implementation details}

\subsection{Portable UDP Networking}

There's few portable UDP networking packages in Common Lisp community,
partly because UDP applications is rare. One UNIX derived systems, the
\textsl{IOlib}\footnote{\texttt{http://common-lisp.net/project/iolib}}
package is a good choice for portable networking: it exports the POSIX
compatibility layer through
\textsl{CFFI}\footnote{\texttt{http://common-lisp.net/project/cffi}}
and has a high-level networking package (net.sockets) and a I/O
multiplex package (io.multiplex). However, due to its heavily dependence
on foreign function interface (FFI) and C code, it will be a bit hard
to deliver applications into single standalone executions on commercial
Common Lisp platforms such as LispWorks. After some investigation,
the \textsc{usocket}\footnote{\texttt{http://common-lisp.net/project/usocket}}
project was been chosen to extend the support on UDP/IP, because \textsc{usocket}
already has a very nice networking API framework.

\textsc{usocket} is much simpler than \textsl{IOlib}. It tries to use
networking APIs which each supported CL implementations already have, and
add foreign functions (as Lisp code) through FFI interface of their own when
necessary. So there's no dependency on \textsl{CFFI} and any other C
code (except on ECL, its FFI interface need C code as embeddable). The
\textsc{usocket} project also has a high-level \texttt{wait-for-input}
function which work in front of UNIX system call \texttt{select()}
or other similar funtions, so users can use \texttt{wait-for-input)}
to swap multiple UDP messages from multiple sockets simultaneously
in one thread.

An \textsc{usocket-udp}
\footnote{\texttt{http://common-lisp.net/projects/cl-net-snmp/usocket.html}} sub-project has been written for the SNMP package,
it implements additional API which is suggested by Erik Huelsmann,
the \textsc{usocket} maintainer. The new functions \texttt{socket-send} and
\texttt{socket-receive} can be used to operate on a new class of
socket called \texttt{datagram-usocket}. \textsc{usocket-udp} depends
on \textsc{usocket} 0.4.x, with the first 0.4.0 released on Oct 28,
2008. Erik also accepted me to the \textsc{usocket} team, which the
next major release will contain the UDP support.

Another issue in UDP network programming is that user code may deal
with packet loss, because UDP is not reliable. A simple way to handle
it is used by Net-SNMP project: define a maximum retry time
and a timeout value, and resend messages on timeout. \textsc{cl-net-snmp}
adopted a more complicated model, it used an ``auto retransmit''
approach \cite{Jacobson:RTT} which usually used in TCP networking: the
timeout value is not fixed but calculated by actual message round-trip
time (RTT). The maximum retry time is still a fix number, as the
timeout value being a range (default is 2~60 seconds). A new
high-level \texttt{socket-sync} function has been defined to do this
automatically.

\subsection{ASN.1 to Common Lisp language mapping}

ASN.1 (Abstract Syntax Notation One) \cite{ISO:ASN.1} is an
international standard which aims at specifying data used in
telecommunication protocols. For more details on ASN.1 and its
history, see Olivier Dubuisson's famous book \cite{Book:ASN.1}.

SNMP highly depends on ASN.1: SNMP MIB (Management Information Base)
\cite{RFC:3418} is full defined in SMI (Structure Management Information)
language, a subset of ASN.1; The most basic data type in SNMP, object identifier (OID),
is just a standard ASN.1 type; all data in SNMP message are enclosed as an ASN.1 \texttt{SEQUENCE} which is then
encoded by BER (Basic Encode Rule) as one of
encoding/decoding methods for ASN.1.

In ASN.1 package, the ASN.1 \texttt{SEQUENCE} type is generally mapped
to Common Lisp \texttt{sequence}, which has two subtypes: \texttt{vector}
and \texttt{list}.
There's only one exception: the empty ASN.1 \texttt{SEQUENCE} is
mapped into empty vector \texttt{\#()} instead of \texttt{nil}, the empty
list. That's because \texttt{nil} is already mapped to ASN.1 type \texttt{NULL},
which is also the only valid element of this type.

There're other ASN.1 primitive types such as all kinds of strings and numbers
which are used by SNMP and they're mapped into correspond Common Lisp types.
Table \ref{table:asn.1-type-mapping} shows most of these type mapping
the ASN.1 package currently supports. Some ASN.1 types are mapped
into CLOS classes.\footnote{In \textsc{cl-net-snmp} 5.x, strings and integers
are just mapped into CL types \texttt{string} and \texttt{integer}. To support
SMI textual conventions (TC, see \cite{RFC:2579}), more complex mapping is needed.
This will be done in next major \textsc{cl-net-snmp} version.}

\begin{table}
  \centering
  \caption{ASN.1 to Common Lisp Type Mapping}
  \label{table:asn.1-type-mapping}
  \begin{tabular}{|l|l|}
    \hline
    \textbf{ASN.1 Type} & \textbf{Common Lisp Type}\\
    \hline
    \texttt{OBJECT IDENTIFIER} & \texttt{ASN.1:OBJECT-ID}\\
    \texttt{INTEGER} & \texttt{CL:INTEGER}\\
    \texttt{NULL} & \texttt{CL:NULL}\\
    \texttt{SEQUENCE} & \texttt{CL:SEQUENCE}\\
    \texttt{OCTET STRING} & \texttt{CL:STRING}\\
    \texttt{IPADDRESS} & \texttt{ASN.1:IPADDRESS}\\
    \texttt{COUNTER32} & \texttt{ASN.1:COUNTER32}\\
    \texttt{COUNTER64} & \texttt{ASN.1:COUNTER64}\\
    \texttt{GAUGE} & \texttt{ASN.1:GAUGE}\\
    \texttt{TIMETICKS} & \texttt{ASN.1:TIMETICKS}\\
    \texttt{OPAQUE} & \texttt{ASN.1:OPAQUE}\\
    \hline
  \end{tabular}
\end{table}

In current released \textsc{cl-net-snmp} versions, ASN.1 module is not
supported. That is: all MIB definitions are compiled into Common Lisp
code in package \texttt{ASN.1}. This may cause symbol clash. Recently this
issue has been solved, now ASN.1 modules are directly mapped into
Common Lisp packages as their original module names.

\subsubsection{BER support}

The BER (Basic Encoding Rule) \cite{ISO:BER} is
essential to implement SNMP because it makes the connection between
ASN.1 object and actual networking packets. BER encodes
every ASN.1 object into three parts: type, length and value (TLV).
The corresponding API funtions for
BER support in ASN.1 package are \texttt{asn.1:ber-encode} and
\texttt{asn.1:ber-decode}. The function \texttt{asn.1:ber-encode} accepts
any Lisp object and try to encode it into a vector of octets
according to BER, and \texttt{asn.1:ber-decode} accepts
a sequence of octets or any octet stream and try to decode into
correspond Lisp object. For example, an integer 10000 can be
encoded into four bytes: 2, 2, 39 and 16, of which the first ``2''
means ASN.1 type \texttt{INTEGER}, the second ``2'' means following
part has \textsl{two} bytes, and 39 and 16 mean the actual value
is 10000 ($39*256+16 = 10000$):

\begin{verbatim}
> (asn.1:ber-encode 10000)
#(2 2 39 16)

> (asn.1:ber-decode #(2 2 39 16))
10000
\end{verbatim}

Another typical example is the encoding of an ASN.1 \texttt{SEQUENCE}.
This type is usually used to implement structure in other languages.
The elements of an ASN.1 \texttt{SEQUENCE} can be anything include
\texttt{SEQUENCE}.  For example, a sequence which contains another
sequence which contains an integer 100, a string ``abc'', and a
\texttt{NULL} data can be expressed into \texttt{\#(\#(100 "abc"
  nil))} in Common Lisp according to our language mapping design. It
can be encoded and decoded correctly:
\begin{verbatim}
> (asn.1:ber-encode #(#(100 "abc" nil)))
#(48 12 48 10 2 1 100 4 3 97 98 99 5 0)

> (asn.1:ber-decode *)
#(#(100 "abc" NIL))
\end{verbatim}

The type byte of sequence is 48. Three elements in inner sequence can
be seen as encoded bytes: \texttt{\#(2 1 100)} (integer 100),
\texttt{\#(4 3 97 98 99)} (string \texttt{"abc"}), and \texttt{\#(5 0)}
(\texttt{nil}).

Both \texttt{ASN.1:BER-ENCODE} and \texttt{ASN.1:BER-DECODE} are
CLOS-based generic functions. \texttt{ASN.1:BER-ENCODE} dispatches on
Common Lisp types, for example the \texttt{INTEGER}:

\begin{verbatim}
(defmethod ber-encode ((value integer))
  (multiple-value-bind (v l)
      (ber-encode-integer value)
    (concatenate 'vector
                 (ber-encode-type 0 0 +asn-integer+)
                 (ber-encode-length l)
                 v)))
\end{verbatim}

The method \texttt{(METHOD ASN.1:BER-ENCODE (INTEGER))} generates a
vector containing type bytes, length bytes and encoding bytes of the
integer. When decoding on integers, generic function
\texttt{ASN.1:BER-DECODE} accepts sequences or streams which contain
data, and then call \texttt{ASN.1:BER-DECODE-VALUE} which dispatches
on keywords (\texttt{:integer} here):

\begin{verbatim}
(defmethod ber-decode ((value sequence))
  (let ((stream (make-instance 'ber-stream
                               :sequence value)))
    (ber-decode stream)))

(defmethod ber-decode ((stream stream))
  (multiple-value-bind (type type-bytes)
      (ber-decode-type stream)
    (multiple-value-bind (length length-bytes)
        (ber-decode-length stream)
      (if type
        (ber-decode-value stream type length)
        ;; When unknown type detected, recover the
        ;; whole data into an ASN.1 RAW object.
        (ber-decode-value
          stream type
          (cons length
            (append type-bytes length-bytes)))))))

(defmethod ber-decode-value ((stream stream)
                             (type (eql :integer))
                             length)
  (declare (type fixnum length) (ignore type))
  (ber-decode-integer-value stream length))
\end{verbatim}

This BER engine in ASN.1 packages is extensible. That's the biggest
difference from other existing BER engines for Common Lisp which can be
found in \textsc{sysman} and \textsc{trivial-ldap}.  The first
value returned by \texttt{asn.1:ber-decode-type} comes from a
hash-table \texttt{asn.1::*ber-dispatch-table*}, and all ASN.1 types
are registered into this hash-table:

\begin{verbatim}
(defun install-asn.1-type (type class p/c tags)
  (setf (gethash (list class p/c tags)
                 *ber-dispatch-table*)
        type))

(install-asn.1-type :integer 0 0 +asn-integer+)
\end{verbatim}

\subsubsection{MIB support}

The ASN.1 \texttt{OBJECT IDENTIFIER} (OID) type is the most important
type in ASN.1.
The way to handle the structure of OID instances
consist the biggest differences between ASN.1 implementations.
Most implementations store full OID number list in each OID
instance.
In \textsc{cl-net-snmp}, ASN.1 OID type in defined by
\texttt{asn.1:object-id} class. Different with most ASN.1 implementations,
\texttt{asn.1:object-id} instances doesn't hold the full OID number list
but only the last one. To construct
a complete OID number list, the rest information is accessed through
the ``parent'' OID instance of the current one. The definition of the
class \texttt{asn.1:object-id} is shown in figure \ref{defclass:object-id}.

\begin{figure}
\begin{verbatim}
(defclass object-id (asn.1-type)
  ((name        :type symbol
                :reader oid-name
                :initarg :name)
   (value       :type integer
                :reader oid-value
                :initarg :value)
   (type        :type oid-type
                :reader oid-type
                :initarg :type)
   (syntax      :type oid-syntax
                :reader oid-syntax
                :initarg :syntax)
   (max-access  :type access
                :reader oid-max-access
                :initarg :max-access)
   (status      :type status
                :reader oid-status
                :initarg :status)
   (description :type string
                :reader oid-description
                :initarg :description)
   (module      :type symbol
                :reader oid-module
                :initarg :module)
   (parent      :type object-id
                :reader oid-parent
                :initarg :parent)
   (children    :type hash-table
                :accessor oid-children
                :initform (make-hash-table)))
  (:documentation "OBJECT IDENTIFIER"))
\end{verbatim}
  \caption{The definition of \texttt{asn.1:object-id} class}
  \label{defclass:object-id}
\end{figure}

The only necessary slots are \texttt{parent} and \texttt{value}.
The \texttt{name} slot is only used by named OID instances (OIDs defined in MIB).
The interface function to build or access an OID instance is
\texttt{ASN.1:OID}.  For example, the OID ``sysDescr.0''
(1.3.6.1.2.1.1.1.0) may be accessed through many ways, which is shown
in Table \ref{table:oid}.

\begin{table}
  \centering
  \caption{Different ways to represent OID ``sysDescr.0''}
  \label{table:oid}
  \begin{tabular}{|l|}
    \hline
    \texttt{(asn.1:oid "sysDescr.0")}\\\hline
	\texttt{(asn.1:oid "SNMPv2-MIB::sysDescr.0")}\\\hline
    \texttt{(asn.1:oid "system.sysDescr.0")}\\\hline
    \texttt{(asn.1:oid "1.3.6.1.2.1.1.1.0")}\\\hline
    \texttt{(asn.1:oid ".1.3.6.1.2.1.1.1.0")}\\\hline
    \texttt{(asn.1:oid "0.1.3.6.1.2.1.1.1.0")}\\\hline
    \texttt{(asn.1:oid \#(1 3 6 1 2 1 1 1 0))}\\\hline
    \texttt{(asn.1:oid '(1 3 6 1 2 1 1 1 0))}\\\hline
    \texttt{(asn.1:oid (list (asn.1:oid "sysDescr") 0))}\\\hline
    \texttt{(asn.1:oid (list |SNMPv2-MIB|::|sysDescr| 0))}\\
    \hline
  \end{tabular}
\end{table}

The MIB node name ``sysDescr'' is pre-defined in Lisp code which is
generated from its MIB definitions by the ASN.1 compiler. Almost all MIB
files shipped with Net-SNMP are provided by the SNMP package.

In recent \textsc{cl-net-snmp}, ASN.1 modules has been mapped
into Common Lisp packages. Each named OID is actually a Lisp variable
in their correspond package, the ``sysDescr'' OID instance is stored
in \texttt{|SNMPv2-MIB|::|sysDescr|}:
\begin{verbatim}
> |SNMPv2-MIB|::|sysDescr|
#<ASN.1:OBJECT-ID SNMPv2-MIB::sysDescr (1) [0]>
\end{verbatim}
The internal structure of this OID instance is shown in Table
\ref{table:object-id}.

\begin{table}
  \centering
  \caption{Internal structure of \texttt{OBJECT-ID} instance}
  \label{table:object-id}
  \begin{tabular}{|l|l|}
    \hline
    \textbf{Slot} & \textbf{Value}\\
    \hline
    \texttt{children} & \texttt{\#<EQL Hash Table\{0\} 21B1550B>}\\
    \texttt{description} & \texttt{"A textual description of the entity."}\\
    \texttt{max-access} & \texttt{ASN.1::|read-only|}\\
    \texttt{module} & \texttt{ASN.1::|SNMPv2-MIB|}\\
    \texttt{name} & \texttt{|ASN.1/SNMPv2-MIB|::|sysDescr|}\\
    \texttt{parent} & \texttt{\#<ASN.1:OBJECT-ID}\\
    & \texttt{\ \ SNMPv2-MIB::system (1) [9]>}\\
    \texttt{status} & \texttt{ASN.1::|current|}\\
    \texttt{syntax} & \texttt{T}\\
    \texttt{type} & \texttt{ASN.1::OBJECT-TYPE}\\
    \texttt{value} & \texttt{1}\\
    \hline
  \end{tabular}
\end{table}

Compared with its original definition in ``SNMPv2-MIB.txt'', almost
all information except the SYNTAX part is saved:

\begin{verbatim}
sysDescr OBJECT-TYPE
    SYNTAX      DisplayString (SIZE (0..255))
    MAX-ACCESS  read-only
    STATUS      current
    DESCRIPTION
        "A textual description of the entity.
         This value should include the full
         name and version identification of
         the system's hardware type, software
         operating-system,
         and networking software."
    ::= { system 1 }
\end{verbatim}

In Table \ref{table:object-id}, the value of slot \texttt{parent}
is another OID instance, which is stored in variable
\texttt{|SNMPv2-MIB|::|system|}:

\begin{verbatim}
> (asn.1:oid-parent |SNMPv2-MIB|::|sysDescr|)
#<ASN.1:OBJECT-ID SNMPv2-MIB::system (1) [9]>

> |SNMPv2-MIB|::|system|
#<ASN.1:OBJECT-ID SNMPv2-MIB::system (1) [9]>

> (eq * **)
T

> (asn.1:oid-name-list |SNMPv2-MIB|::|system|)
("iso" "org" "dod" "internet" "mgmt" "mib-2"
 "system")

> (asn.1:oid-number-list |SNMPv2-MIB|::|system|)
(1 3 6 1 2 1 1)
\end{verbatim}

Actually all named OID instances have another
named OID instances as their parent or children which can be
accessed from their corresponding slots; all and only these named
OID instances are stored as one conceptual ``MIB tree''.
OID instances which doesn't have a name (like
``sysDescr.0'') are created by \texttt{ASN.1:OID} function when it's been called every time. For these unnamed OID instances, the \texttt{parent} slot are
used for them to track back full OID number list when being used by SNMP operations.

The ``root'' node of the MIB tree is \texttt{(OID "zero")}, which
is also assigned in Lisp variable \texttt{ASN.1::*ROOT-OBJECT-ID*}.
Its the entry to all named MIB nodes:

\begin{verbatim}
> asn.1::*root-object-id*
#<ASN.1:OBJECT-ID zero (0) [2]>

> (asn.1::list-children *)
(#<ASN.1:OBJECT-ID iso (1) [1]>
 #<ASN.1:OBJECT-ID SNMPv2-SMI::zeroDotZero (0) [0]>)
\end{verbatim}

\subsection{SNMP internal}

Encryption and authentication support in \textsl{SNMPv3} need HMAC,
DES, MD5 and SHA1 algorithms. This is already done by Nathan Froyd's \textsc{ironclad}%
\footnote{\texttt{http://method-combination.net/lisp/ironclad/}}
project, which supplies almost all authenticate and encryption
algorithms written in pure Common Lisp.

The internal work of SNMP interface functions is rather
straightforward, following steps will happen:
\begin{enumerate}
\item Prepare a variable bindings list according to SNMP operation type and arguments.
\item Create a \texttt{pdu} instance using above variable bindings list.
\item Create a \texttt{message} instance and put above \texttt{pdu} instance in it.
\item Encode the \texttt{message} to get the sending packet data.
\item Send it, then get a response packet.
\item Decode the response packet and create a new \texttt{message} instance from the decode result and the old \texttt{message} instance.
\item Retrieve the variable bindings list from the \texttt{pdu} slot in above \texttt{message} instances.
\item Generate return values from above variable bindings list.
\end{enumerate}
Though there're many steps on processing SNMP operations
\cite{RFC:3412}, the core function,
\texttt{snmp::snmp-request}, which do most steps in above and it's quite simple.
The source code is shown in Figure \ref{func:snmp-request}.

\begin{figure*}
\begin{verbatim}
(defmethod snmp-request ((session session) (request symbol) (bindings list)
                         &key context)
  (when bindings
    (let ((vb (mapcar #'(lambda (x) (if (consp x)
                                        (list (oid (first x)) (second x))
                                        (list (oid x) nil)))
                      bindings)))
      ;; Get a report first if the session is new created.
      (when (and (= (version-of session) +snmp-version-3+)
                 (need-report-p session))
        (snmp-report session :context context))
      (let ((message (make-instance (gethash (type-of session) *session->message*)
                                    :session session
                                    :context (or context *default-context*)
                                    :pdu (make-instance request
                                                        :variable-bindings vb))))
        (let ((reply (send-snmp-message session message)))
          (when reply
            (map 'list #'(lambda (x) (coerce x 'list))
                 (variable-bindings-of (pdu-of reply)))))))))

(defun snmp-get (session bindings &key context)
  (let ((result (mapcar #'second
                        (snmp-request session 'get-request-pdu bindings
                                      :context context))))
    (if (consp bindings) result (car result))))
\end{verbatim}
  \caption{The core SNMP function: \texttt{snmp-request}}
  \label{func:snmp-request}
\end{figure*}

\section{Future Work}

There's still lots of work to do. On ASN.1 side, the SMIv2
Textual Convention (TC) \cite{RFC:2579} haven't been implemented.
This part of work
will give a better representation on strings and numbers used in
SNMP.

On client side SNMP, to fulfill the high performance requirements
of enterprise applications, the SNMP client must be able to do
multiple SNMP operations at the same
time in a single thread, and even using sockets less than the number
of remote SNMP peers. This feature has been asked by some customers,
and there's already a MSI project\footnote{
\texttt{http://www.msi.co.jp/\~{}fremlin/projects/snmp-async/}} which try
to implement this feature on top of \textsc{cl-net-snmp}.

On server side SNMP, the VACM (View-based Access
Control Model) \cite{RFC:3415} is on the top of the TODO
list, and it will be implemented in next \textsc{cl-net-snmp} version.

There're also plans on improving GUI and Web interface of the SNMP package.
Currently the GUI interface has only a graphical MIB browser based on
LispWorks CAPI toolkit\footnote{
\texttt{http://www.lispworks.com/products/capi.html}}, and it will be
extend to a full featured SNMP GUI client tool and maybe turn to
support the Common Lisp Interface Manager (CLIM). The Web interface
will be based on CL-HTTP and try to provide a HTTP interface for client and 
server side SNMP work.

Obviously SNMP is not the only networking protocol which is based on
ASN.1 and UDP.  The existing work of \textsc{cl-net-snmp} could be
used to develop Common Lisp packages of other related networking
protocols.

Lightweight Directory Access Protocol (LDAP) \cite{RFC:4510}
is just another important
protocol which is completely based on ASN.1. LDAP is widely used for
management in large busyness and there're many ``directory server''
productions. Currently the LDAP support in Common Lisp is still in its
early stage, only a few of small packages being developed. A new LDAP
package based on ASN.1 package is on the plan list of
\textsc{cl-net-snmp} project.

Intelligent Platform Management Interface (IPMI) \footnote{
\texttt{http://www.intel.com/design/servers/ipmi/}} is also
another network management protocol beside SNMP. It's a UDP-based
protocol which is slightly easier than SNMP and usually implemented
directly by server hardware. Through IPMI, system administrators can
power off a remote server, query its hardware log, or logging to the
server console through the Serial-On-LAN (SOL) interface. An IPMI
package based on existing portable UDP networking package is in
progress.

\acks

This project is part of a research project on network management platform
which supported by NetEase.com, Inc.

Thanks will go to the following Common Lisp packages and
implementations which are depended on by \textsc{cl-net-snmp} project
and their major author:
\begin{itemize}
\item \textsc{sysman} Project (Simon Leinen),
\item \textsc{usocket} Project (Erik Huelsmann),
\item GBBopen Project\footnote{\texttt{http://gbbopen.org}} (Daniel D. Corkill),
\item \textsc{ironclad} Project (Nathan Froyd),
\item LispWorks\footnote{\texttt{http://www.lispworks.com}}.
\end{itemize}

Special thanks to Mathematical Systems Inc. (MSI) \footnote{
\texttt{http://www.msi.co.jp}} for the adoption of
\textsc{cl-net-snmp} in their enterprise projects,
with bugfix and new feature suggestion.

\bibliographystyle{plainnat}

\end{document}